\documentclass[%
 prl,
rsi,%
 amsmath,amssymb,
 reprint,%
]{revtex4-1}

\usepackage{graphicx}
\usepackage{dcolumn}
\usepackage{bm}
\usepackage{dsfont}

\graphicspath{
	{figures/}
}

\begin{document}

\title{Core-valence attosecond transient absorption spectroscopy of polyatomic molecules}%

\author{Nikolay V. Golubev}
\email{nik.v.golubev@gmail.com}
\affiliation{Laboratory of Theoretical Physical Chemistry, Institut des Sciences et Ing\'enierie Chimiques, Ecole Polytechnique F\'ed\'erale de Lausanne (EPFL), CH-1015 Lausanne, Switzerland}

\author{Ji\v{r}\'{\i} Van\'{\i}\v{c}ek}
\affiliation{Laboratory of Theoretical Physical Chemistry, Institut des Sciences et Ing\'enierie Chimiques, Ecole Polytechnique F\'ed\'erale de Lausanne (EPFL), CH-1015 Lausanne, Switzerland}

\author{Alexander I. \surname{Kuleff}}
\affiliation{Theoretische Chemie, Universit\"at Heidelberg, Im Neuenheimer Feld 229, D-69120 Heidelberg, Germany}
\affiliation{ELI-ALPS, Wolfgang Sandner utca 3, H-6728 Szeged, Hungary}

\date{\today}

\begin{abstract}
Tracing ultrafast processes induced by interaction of light with matter is often very challenging. In molecular systems, the initially created electronic coherence becomes damped by the slow nuclear rearrangement on a femtosecond timescale which makes real-time observations of electron dynamics in molecules particularly difficult. In this work, we report an extension of the theory underlying the attosecond transient absorption spectroscopy (ATAS) for the case of molecules, including a full account for the coupled electron-nuclear dynamics in the initially created wave packet, and apply it to probe the oscillations of the positive charge created after outer-valence ionization of the propiolic acid molecule. By taking advantage of element-specific core-to-valence transitions induced by X-ray radiation, we show that the resolution of ATAS makes it possible to trace the dynamics of electron density with atomic spatial resolution.
\end{abstract}

\pacs{Valid PACS appear here}
\maketitle

The impressive progress in laser technologies during last few decades~\cite{krausz2009,calegari2016} has stimulated the rapid development of atomic and molecular physics. With the advent of ultrashort laser pulses, the scientific community obtained a unique tool to study such phenomena as charge transport in molecules~\cite{lewis2012,worner2017_chmig_rev} and elementary steps of chemical reactions~\cite{zewail2000d,baker2006}, to name a few. Being able to observe fundamental processes with attosecond temporal resolution, it becomes also possible to steer and probe electron dynamics on its natural time scale~\cite{ramasesha2016}.

One phenomenon of particular interest for the attosecond science is the process of ultrafast charge migration driven solely by the electron correlation and electron relaxation~\cite{cederbaum1999}. Charge migration arises whenever a coherent superposition of multiple electronic states is prepared, which, due to the electron correlation, can be achieved even when one removes an electron from a single molecular orbital. Such a coherent population of states can thus result from ionization of both outer- and inner-valence shells of a molecule~\cite{breidbach2003,lunnemann2008a,lunnemann2009}, and even by removing an electron from localized core orbitals~\cite{kuleff2016}. The dynamics triggered by the ionization manifests as a migration of the initially created localized charge through the system.

Although much had been achieved in the theoretical understanding of the ultrafast electronic processes taking place in molecules~\cite{kuleff2014,nisoli2017}, the experimental demonstration of the electronic dynamics is still in its infancy~\cite{leone2014}. So far, only a very limited number of experimental studies have been performed that were able to explore the correlated electron motion in complex systems. Evidence of ultrafast charge migration in amino acids phenylalanine~\cite{belshaw2012,calegari2014} and tryptophan~\cite{lara-astiaso2018} was demonstrated by measuring the yield of a doubly charged ion as a function of the delay between the ionizing pump and doubly ionizing probe pulses.

An alternative technique, using a rescattering of the ionized electron as a probe, is to measure the resulting high-harmonic generation (HHG) spectra~\cite{smirnova2009,mairesse2010}. In the seminal work by Kraus \textit{et al.}~\cite{kraus2015}, attosecond charge migration in ionized iodoacetylene was reconstructed and controlled by analyzing the HHG spectrum emitted after irradiation of the molecule with strong infrared pulses of different wavelength (800 and 1300~nm). Despite being a promising experimental technique, the HHG spectroscopy can currently be used to capture only the first few femtoseconds of the dynamics thus preventing a simultaneous observation of the initial coherent electron motion and the follow-up effects caused by the coupling between the fast moving electrons and the slower nuclei.

Another experimental approach which combines high spectral and high temporal resolution is the attosecond transient absorption spectroscopy (ATAS)~\cite{goulielmakis2010,santra2011,wirth2013}. Measuring the transmission/absorption of a broadband laser pulse through a sample, one can gain intuitive and highly detailed insights about dynamics of the system. In the pioneering work by Goulielmakis \textit{et al.}~\cite{goulielmakis2010}, the ATAS was used to trace in real-time the valence electron motion in strong-field-generated Kr$^+$ ions. The straightforward interpretation of the ATAS~\cite{santra2011} makes this technique a promising tool to study ultrafast dynamics in more complex molecular systems~\cite{drescher2019,kobayashi2019,kobayashi2020,kobayashi2020a,timmers2019}.

A particularly appealing feature of ATAS is the possibility to trace the evolution of the system with atomic spatial resolution by probing electron motion at a specific site of the molecule, as originally proposed by Dutoi \textit{et al.}~\cite{dutoi2013,dutoi2014}. The latter is possible by taking advantage of element-specific core-to-valence transitions induced by X-ray radiation. Recently, experimental setups providing soft X-ray pulses with sub-femtosecond duration were reported in both table-top~\cite{saito2019,barreau2020} and free electron laser facilities~\cite{duris2020}. ATAS, therefore, appears as a very promising technique for the investigation of the charge migration dynamics triggered by ionization~\cite{li2020}.

So far, the theory underlying ATAS has only been developed for atoms~\cite{santra2011,pabst2012,kolbasova2019}, or for molecules either in absence of nuclear motion~\cite{hollstein2017,drescher2019} or in absence of electronic coherence \cite{kobayashi2019}. At the same time, extensive \textit{ab initio} calculations for several molecules demonstrated~\cite{arnold2017,vacher2017,despre2018,golubev2020TGA} that coupling between electronic and nuclear degrees of freedom has a dramatic impact on the electron dynamics and thus has to be taken into account when computing the transient absorption spectrum. While the extensions of the ATAS formalism to the case of molecules~\cite{baekhoj2016,rorstad2018} have been widely used to support recent experimental measurements (see, e.g., Refs.~\cite{liao2017,saito2019}), the reported schemes are all based on the numerical analysis of various correlation functions which are difficult to interpret and numerically demanding to compute. In contrast, following the procedure reported in Ref.~\cite{santra2011} for the case of atoms, we present a simple quasi-analytical expression for the absorption cross-section of molecules which accounts for the nuclear motion and non-adiabatic dynamics and is composed from physically intuitive terms.

In this Letter, we extend the theory of ATAS to the case of polyatomic molecules, fully accounting for the coupled electron-nuclear dynamics in the initially created wave packet. We demonstrate on the example of propiolic acid that the resolution of ATAS is sufficient to clearly observe the ultrafast oscillations of the electron density, as well as the follow-up decoherence caused by the nuclear rearrangement. We perform high-level \textit{ab initio} simulations of X-ray ATAS, showing the possibility to trace the dynamics of the created hole through the molecular chain, and thus demonstrate that X-ray ATAS can be a very useful technique to study ultrafast electron motion in molecules.

Ionization of a molecule by an ultrashort laser pulse brings the system to a non-stationary superposition of ionic states
\begin{equation}
\label{eq:full_WF}
	\Psi(\mathbf{r},\mathbf{R},t) = \sum_I \chi_I(\mathbf{R},t) \Phi_I (\mathbf{r},\mathbf{R}),
\end{equation}
where $\mathbf{r}$ and $\mathbf{R}$ denote electronic and nuclear coordinates, respectively, $\Phi_I (\mathbf{r},\mathbf{R})$ are the electronic eigenstates obtained by solving the stationary Schr\"odinger equation $\hat{H}_e \Phi_I (\mathbf{r},\mathbf{R}) = E_I(\mathbf{R}) \Phi_I (\mathbf{r},\mathbf{R})$ with electronic Hamiltonian $\hat{H}_e$, and $\chi_I(\mathbf{R},t)$ are the expansion coefficients representing nuclear wave packets moving on the corresponding potential energy surfaces (PESs) $E_I(\mathbf{R})$. The ansatz~(\ref{eq:full_WF}) for the full molecular wavefunction is, in principle, an exact way to describe the concerted motion of electrons and nuclei in a molecule.

The electronic and nuclear dynamics induced after ionization of a system can be probed by analyzing the absorption of a spectrally broadband short laser pulse. Employing the time-dependent perturbation theory and Condon approximation, the absorption cross-section can be calculated as
\begin{widetext}
\begin{equation}
\label{eq:cross_section}
	\sigma(\omega,\tau) = \frac{4 \pi \omega}{c} \text{Im} 
		\sum_I \sum_J \langle \chi_I (\mathbf{R},\tau) | \chi_J (\mathbf{R},\tau) \rangle_{\mathbf{R}}
		\sum_F \langle \Phi_I | \hat{\mu} | \Phi_F \rangle \langle \Phi_F | \hat{\mu} | \Phi_J \rangle
		\left(
			\frac{1}{\tilde{E}_F - E_I - \omega} + \frac{1}{\tilde{E}_F^* - E_J + \omega}
		\right),
\end{equation}
\end{widetext}
where $\omega$ is the photon energy, $\tau$ is the delay between pump and probe pulses, $c$ is the speed of light in vacuum, $\langle \Phi_I | \hat{\mu} | \Phi_F \rangle$ and $\langle \Phi_F | \hat{\mu} | \Phi_J \rangle$ denote transition dipole matrix elements between initial $\{I,J\}$ and final $F$ electronic states with corresponding energies $E_{\{I,J,F\}}$ computed at fixed geometry $\mathbf{R}_0$, and quantities $\langle \chi_I (\mathbf{R},\tau) | \chi_J (\mathbf{R},\tau) \rangle_{\mathbf{R}}$ represent populations of initial electronic states when $I=J$ and the electronic coherences when $I \neq J$. To account for the broadening of the spectrum, we assign the final states a finite lifetime $1/\Gamma$, which enters Eq.~(\ref{eq:cross_section}) in form of complex final state energies $\tilde{E}_F = E_F - i(\Gamma/2)$. The detailed derivation of Eq.~(\ref{eq:cross_section}) is presented in Sec.~I of the Supplemental Material~\cite{SM}.

As one can see from Eq.~(\ref{eq:cross_section}), the signal represents a combination of contributions corresponding to $\Lambda$-type transitions which couple initially populated $I$ and $J$ states through intermediate excited states $F$. The possibility to measure electron dynamics between states $I$ and $J$ is based on the fact that the interference between absorption of a photon with energy $\omega_1$ and stimulated emission of a photon with a different energy $\omega_2$ takes place only if the initial states are populated coherently. The electronic coherence term $\langle \chi_I (\mathbf{R},\tau) | \chi_J (\mathbf{R},\tau) \rangle_{\mathbf{R}}$ is the only source of time dependence in Eq.~(\ref{eq:cross_section}). Therefore, all the dynamical information required for calculations of absorption cross-section comes from the propagation of nuclear wave packets $\chi_I (\mathbf{R},t)$ in the initially populated electronic states while final states define energy positions and intensities of the corresponding lines in the spectrum.

Numerous theoretical approaches for computing electronic coherences were developed in the last few years~\cite{vacher2015decoh,arnold2017,vacher2017,despre2018,golubev2020TGA}. Here, we use the recently described~\cite{golubev2020TGA} semiclassical thawed Gaussian approximation (TGA) scheme~\cite{heller1975,begusic2018,begusic2019,lasser2020}, which combines efficient single trajectory evaluation of nuclear dynamics with fully \textit{ab initio} on-the-fly simulations of electronic structure. Within the TGA, the nuclear wavefunction is described by a single Gaussian wave packet whose center follows Hamilton's equations of motion and whose time-dependent width and phase are propagated using the local harmonic approximation of the PES. Despite its simplicity, the TGA gives results comparable in accuracy with the full dimensional quantum calculations when the involved states are not strongly coupled by non-adiabatic effects, as explicitly shown for propiolic acid~\cite{golubev2020TGA}.

To describe the electronic states of the ionized system, we use the high-level \textit{ab initio} algebraic diagrammatic construction (ADC) scheme~\cite{schirmer1983} for representing the one-particle Green's function. The ionization out of the valence orbitals is computed at the third-order ADC [ADC(3)]~\cite{schirmer1998}, while core ionic states are modeled using the fourth-order scheme [ADC(4)]~\cite{angonoa1987}. Separate treatment of valence and core orbitals is dictated by the fact that the numerical efforts required for constructing and diagonalizing the ADC(4) secular matrix are considerable and prevents for the moment a straightforward application to the valence ionization of all but the smallest systems. At the same time, accurate description of core electrons requires an explicit account for the large relaxation effects while the correlation between the core and the valence electrons can be neglected. The ADC(4) scheme has been shown many times to yield highly accurate results when applied to core ionization~\cite{feyer2009}. Standard double-zeta plus polarization (DZP) basis sets~\cite{canalneto2005} were employed to construct the noncorrelated reference states. Ground-state geometry of the neutral molecule was optimized using the Gaussian 16 implementation~\cite{g16} of the density functional theory~\cite{hohenberg1964} with the B3LYP functional~\cite{stephens1994}. 

\begin{figure}
\includegraphics[width=8.5cm]{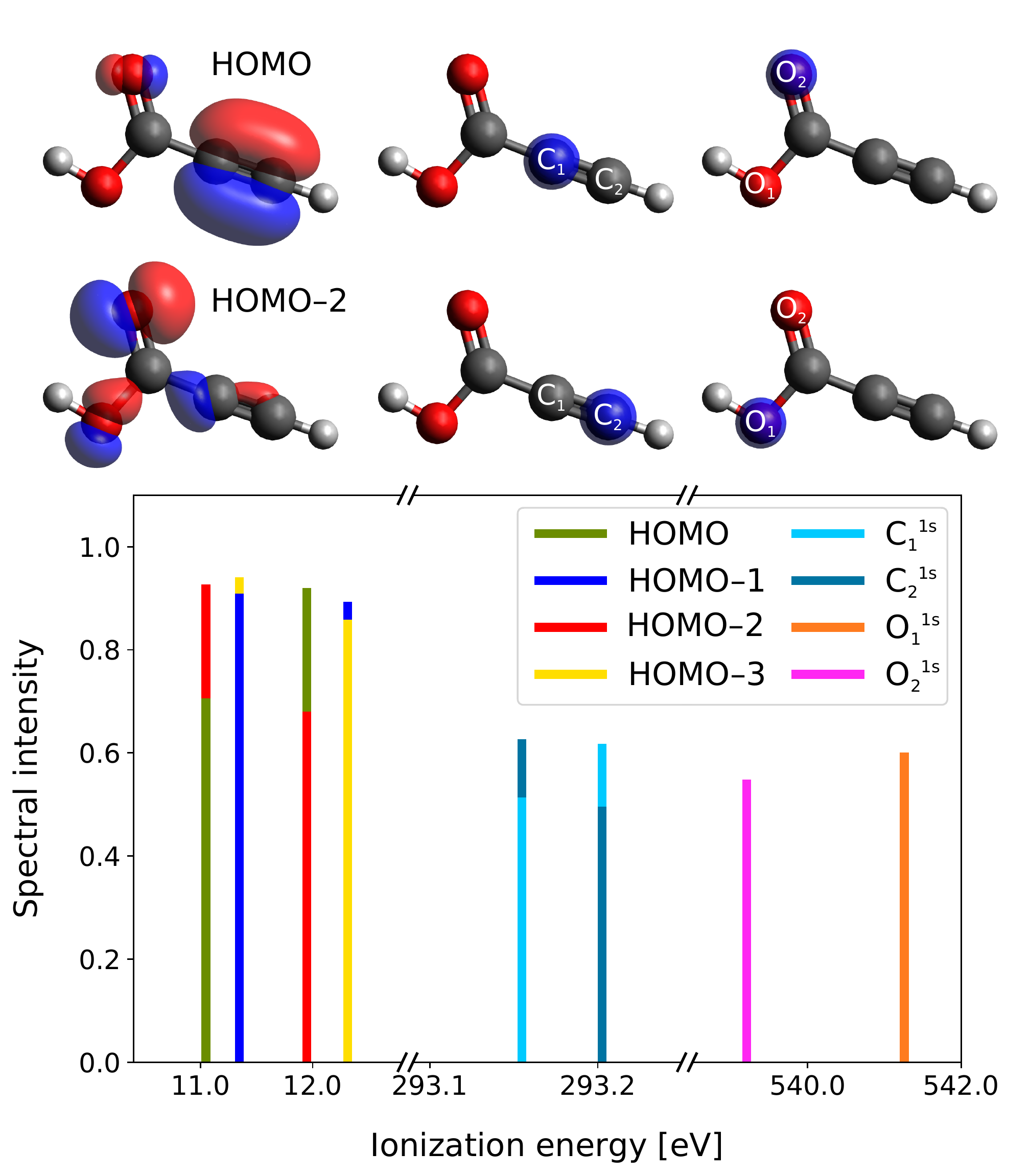}
\caption{Ionization spectrum of propiolic acid computed using the \textit{ab initio} many-body Green's function ADC method. Three energy windows illustrate the valence region ($10.5-12.5$~eV), core ionic states resulting from ionization out of $1s$ orbitals of triple-bond carbon atoms ($293.1-293.25$~eV) and oxygen atoms ($539-542$~eV). In all three energy ranges, the closest ionic states are located at least $\sim$2.5~eV above the presented lines. The spectral intensity is defined as the combined weight of all one-hole configurations in the configuration-interaction expansion of the ionic state. The relevant molecular orbitals are also depicted.}
\label{fig:ionization_prop_acid}
\end{figure}

Ionization spectrum of the propiolic acid is shown in Fig.~\ref{fig:ionization_prop_acid}. The first four valence ionic states are energetically separated by 2.8~eV from the remaining ones. Due to the electron correlation, the ground and the second excited ionic states of the molecule are a strong mixture of two one-hole configurations: an electron missing in the highest occupied molecular orbital (HOMO) and an electron missing in the HOMO-2. Therefore, a sudden removal of an electron either from HOMO or from HOMO-2 will create an electronic wave packet, which will initiate charge migration oscillations between the carbon triple bond and the carbonyl oxygen with a period of about 4.5 fs, determined by the energy gap between the first and the third cationic states~\cite{golubev2017qoc,despre2018}. The two orbitals involved in this hole-mixing are also depicted in Fig.~\ref{fig:ionization_prop_acid}. Importantly, the molecule has planar symmetry and thus belongs to the C$_s$ symmetry group which allows assignment of the ionic states to two irreducible representations: the first and third states belong to the A$'$, while the second and fourth states to A$''$. The latter allows one to obtain the desired superposition of the first and third ionic states by appropriately orienting the molecule with respect to the laser polarization.

Let us now look at the states corresponding to ionization out of the core orbitals of the molecule. Due to the fact that $1s$ core orbitals are localized in space, it is possible to associate every core orbital with a specific atom in the system. Figure~\ref{fig:ionization_prop_acid} depicts ionic states resulting from ionization out of core orbitals of oxygen and carbon atoms of the propiolic acid. The spectrum is plotted for two energy windows which capture ionic states with dominant contributions of the corresponding core orbitals. Interestingly, core orbitals of carbon atoms forming the triple bond experience strong hole mixing similar to the one reported previously for ionization out of valence and upper-range inner orbitals only~\cite{breidbach2003,kuleff2014}. We would like to note in passing that due to the small energy difference between these core ionic states, the electronic oscillations for the wave packet resulting from such a mixture is expected to be significantly slower than those taking place in the case of valence ionization of the molecule. The ionic state resulting from ionization out of $1s$ orbital of carbon atom of the carboxyl group is located 3.9~eV above the states belonging to ionization out of orbitals of carbons forming the triple bond and is not shown in Fig.~\ref{fig:ionization_prop_acid}. Similarly, closest satellite ionic states corresponding to ionization out of oxygen core orbitals are located 2.5~eV above main lines. These energy gaps make it possible to distinguish energetically the transitions between valence ionic states and only those core states shown in Fig.~\ref{fig:ionization_prop_acid} while neglecting transitions to other electronic states which are not covered by the presented energy ranges.

To account for transitions between initial valence $|\Phi_{\{I,J\}} \rangle$ and final core $| \Phi_F \rangle$ states present in Eq.~(\ref{eq:cross_section}), one needs to evaluate the corresponding transition dipole matrix elements $\langle \Phi_I | \hat{\mu} | \Phi_F \rangle$ and $\langle \Phi_F | \hat{\mu} | \Phi_J \rangle$. The latter cannot be done directly, because our initial and final states are computed at different levels of electronic structure theory. An alternative way to perform these calculations is to exploit the configuration interaction (CI)-like structure of the electronic wavefunctions appearing in the ADC approach. Details of this procedure can be found in Sec.~II of~\cite{SM} (see also Ref.~\cite{pohl2017orbkit}). Importantly, accurate values of the dipole transitions play a central role in the possibility to resolve spatial localization of the charge in a molecule. Due to the interference between photons connecting various combinations of initial $\{I,J\}$ and final $F$ states, scaled by the corresponding dipole transitions, one can infer a direct correspondence between the real-space dynamics taking place in a system under study and the ATAS signal computed via Eq.~(\ref{eq:cross_section}).

\begin{figure}
\includegraphics[width=8.5cm]{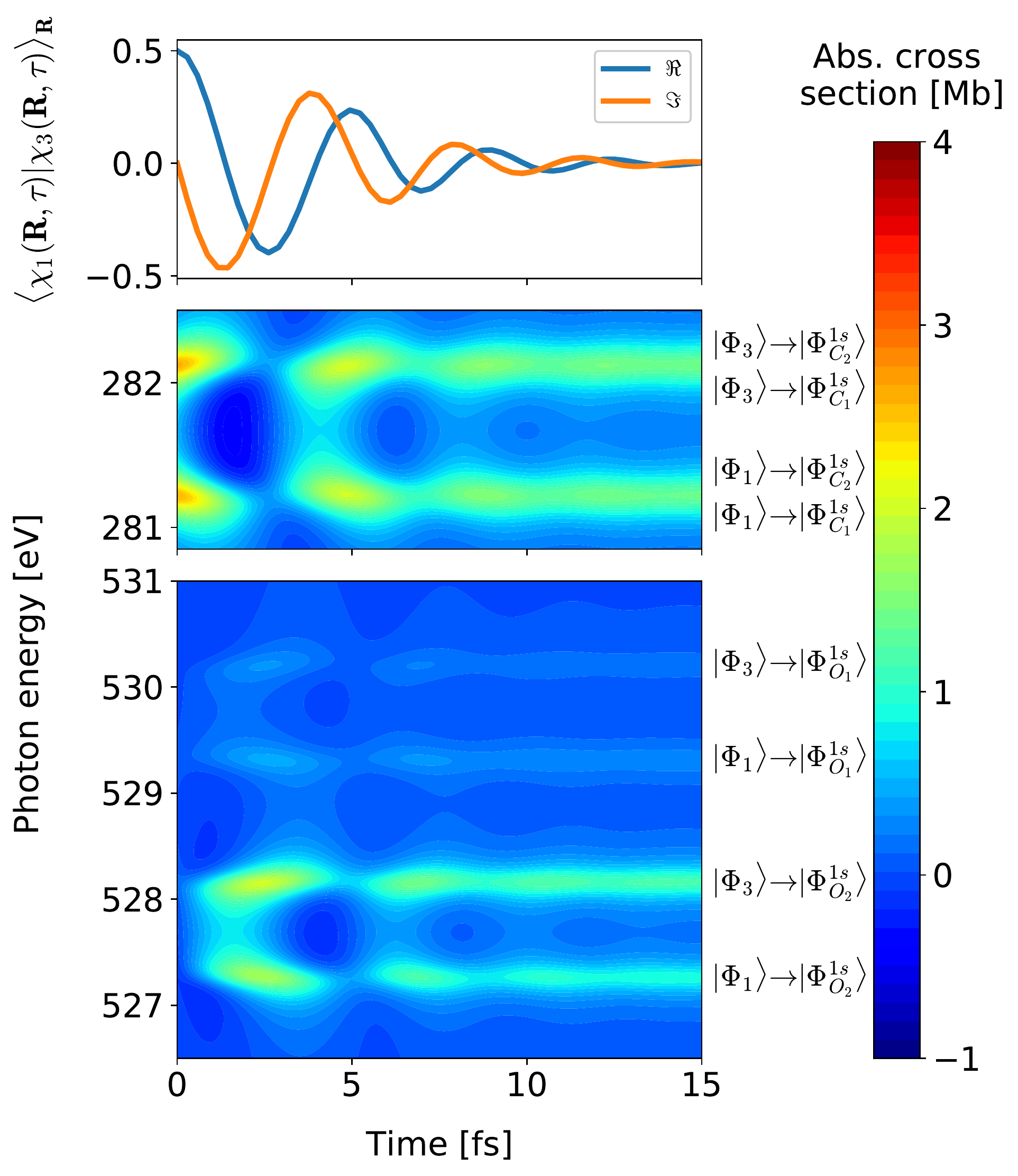}
\caption{Top panel: Electronic coherence measured by the time-dependent overlap $\langle \chi_1 (\mathbf{R},\tau) | \chi_3 (\mathbf{R},\tau) \rangle_{\mathbf{R}}$ of the nuclear wave packets propagated in the first and third cationic states of propiolic acid after the removal of an electron from the HOMO. Middle panel: The time-resolved absorption cross-section as a function of the photon energy and time delay, plotted for the energy window corresponding to transitions between initially populated valence ionic states and the core states resulting from ionization out of $1s$ orbitals of carbon atoms forming the triple bond. Bottom panel: The absorption cross-section plotted for the energy range covering transitions to core ionic states of the oxygen atoms. The ionic states involved in the corresponding transitions are also shown. The transient absorption traces shown are rotationally averaged and thus describe randomly oriented molecules.}
\label{fig:ATAS_prop_acid}
\end{figure}

We performed calculations of ATAS for the evolution of the wave packet created by projecting the ground (electronic and nuclear) neutral state of the propiolic acid onto the first and third cationic states of the molecule employing the sudden and Franck--Condon approximations. The initial phase between involved electronic states is chosen in such a way that the created charge is localized in the HOMO. The top panel of Fig.~\ref{fig:ATAS_prop_acid} shows the evolution of the electronic coherence between initially populated states, i.e. the matrix element $\langle \chi_1 (\mathbf{R},\tau) | \chi_3 (\mathbf{R},\tau) \rangle_{\mathbf{R}}$, in time. It is seen that the electronic oscillations are strongly influenced by the nuclear motion and the coherence is completely suppressed within the first 15~fs~\cite{despre2018,golubev2020TGA}. The middle and bottom panels of Fig.~\ref{fig:ATAS_prop_acid} depict the absorption cross-section computed using Eq.~(\ref{eq:cross_section}) for transitions between initially populated valence ionic states and the core states, resulting from ionization out of $1s$ orbitals of carbon and oxygen atoms, respectively. Line positions in ATAS are determined by the energy gaps between the initial and final electronic states. Transitions from the two initially populated valence states to the two final oxygen core ionic states lead to the appearance of four lines in the spectrum. In the case of carbon core ionic states, the four absorption lines appear as two visible transitions due to the small energy gap between final core states. The intensity of the lines is determined by the product of the transition dipole moments that couple initial and final states via $\Lambda$-type transitions. All calculations were rotationally averaged and thus describe randomly oriented molecules (see Sec.~III of~\cite{SM}). An energy broadening parameter $\Gamma=0.3$~eV was used (see Sec.~I of~\cite{SM}).

Spectra in Fig.~\ref{fig:ATAS_prop_acid} clearly illustrate that the resolution of ATAS makes it possible to trace the real-time oscillatory charge migration dynamics before the nuclear motion dephases the electron coherence. Importantly, the maxima in the beatings of the absorption lines corresponding to core excitations of both carbon and oxygen atoms correlate with the spatial localization of the hole density in the vicinity of these atoms. The latter allows us to trace the electron dynamics along the molecular chain with atomic spatial resolution. It is seen that immediately after the ionization the absorption takes place almost exclusively on the triple-bond carbon atoms suggesting that the initial hole-charge is located in the triple bond of the molecule. Within 2~fs, the charge density migrates to the carboxyl group of the molecule which is reflected by the strong absorption signal from the oxygen atoms, mainly from the carbonyl oxygen. After several oscillations, the lines become stationary yet remain present for all shown core states, which reflects the fact that the charge distributes almost uniformly along the molecular chain.

Before concluding, we emphasize that Eq.~(\ref{eq:cross_section}) is obtained with the assumption that the initial time-dependent state of the system, Eq.~(\ref{eq:full_WF}), is prepared before the action of the probe pulse. In other words, we utilize the first order polarization response of the system with respect to the applied electric field, operating in a regime of non-overlapping pump and probe pulses. Although being a limitation, this treatment of the light-matter interaction allows us to simulate ATAS with the time-dependent perturbation theory applied to a non-stationary initial state of a system that is, in turn, described in a completely nonperturbative fashion. Another major approximation used in this study is the short-pulse approximation for the probe pulse. The latter requirement means that the duration of the probe pulse should be shorter than the oscillations of the coherences in the initially prepared superposition. The reported scheme suggests simultaneous measurements of the absorption cross-section for both carbon and oxygen core ionic states of the propiolic acid which automatically makes the required laser pulse short enough to cover this broad energy range. An ideal experiment would involve ultrashort XUV-pump and X-ray-probe pulses, although the ionization step can be also performed via other mechanisms, e.g., strong-field multiphoton absorption. In the latter case, the initially created superposition can be sensitive to the variations of the laser pulse parameters which will lead to the loss of the coherence and thus to less pronounced ATAS signal (see Sec.~IV of~\cite{SM}). Nevertheless, as was already pointed out in the introduction, modern experimental setups~\cite{barreau2020,duris2020} provide radiation with the unique combination of high intensity, high photon energy, and short pulse duration which can open the door for real-time studies of the electron dynamics in complex molecules using ATAS technique. However, because the electronic states belonging to the ionization out of core orbitals of the same atoms lie typically very close to each other in energy (see, e.g., carbon ionic states in Fig.~\ref{fig:ionization_prop_acid}), the resolution of ATAS might be limited to trace the dynamics only between atoms of different chemical elements, or between differently bonded same elements giving rise to larger chemical shifts.

In conclusion, we have demonstrated the application of ATAS to probe the ultrafast dynamics of electron density in a polyatomic molecule. We presented the extension of the theory underlying ATAS taking into account coupling between electronic and nuclear degrees of freedom. We performed high-level \textit{ab initio} calculations of ATAS for the propiolic acid demonstrating the possibility to detect both temporal and spatial aspects of the electron dynamics. Our findings illustrate that the resolution of ATAS in soft X-ray energy range is sufficient to identify structural groups and even particular atoms involved in the process. Finally, we observed strong hole mixing between core orbitals forming the triple bond of a molecule. The investigation of possible dynamics resulting from ionization out of core orbitals is a promising direction of further research. We hope that our work will motivate such studies.

\begin{acknowledgments}
N.\ V.\ G. acknowledges the support by the Branco Weiss Fellowship---Society in Science, administered by the ETH Z\"urich. J.\ V. acknowledges the Swiss National Science Foundation for financial support through the National Center of Competence in Research MUST (Molecular Ultrafast Science and Technology) Network. A.\ I.\ K. thanks the financial support provided by the DFG through the QUTIF Priority Programme and the US ARO (grant Nr. W911NF-14-1-0383).
\end{acknowledgments}

\bibliographystyle{apsrev4-1}
\bibliography{ATAS_decoherence.bib}

\end{document}


\title{Supplementary Material for ``Core-valence attosecond transient absorption spectroscopy of polyatomic molecules''}%

\author{Nikolay V. Golubev}
\email{nik.v.golubev@gmail.com}
\affiliation{Laboratory of Theoretical Physical Chemistry, Institut des Sciences et Ing\'enierie Chimiques, Ecole Polytechnique F\'ed\'erale de Lausanne (EPFL), CH-1015 Lausanne, Switzerland}

\author{Ji\v{r}\'{\i} Van\'{\i}\v{c}ek}
\affiliation{Laboratory of Theoretical Physical Chemistry, Institut des Sciences et Ing\'enierie Chimiques, Ecole Polytechnique F\'ed\'erale de Lausanne (EPFL), CH-1015 Lausanne, Switzerland}

\author{Alexander I. \surname{Kuleff}}
\affiliation{Theoretische Chemie, Universit\"at Heidelberg, Im Neuenheimer Feld 229, D-69120 Heidelberg, Germany}
\affiliation{ELI-ALPS, Wolfgang Sandner utca 3, H-6728 Szeged, Hungary}

\date{\today}

\begin{abstract}
In the Letter, we present an application of the attosecond transient absorption spectroscopy (ATAS) to probe the electronic oscillations of a positive charge created after outer-valence ionization of the propiolic acid molecule, while accounting for all electronic and nuclear degrees of freedom. Here, we present additional details: in Sec.~I, we derive the expression for the absorption cross-section in ATAS of polyatomic molecules. Section~II contains details of the procedure used for evaluating transition properties between electronic states computed by high-level \textit{ab initio} electronic structure methods. In Secs.~III and IV, we discuss the averaging of the absorption cross-sections over rotational degrees of freedom and with respect to relative contributions of ionic states to the initial wave packet, respectively.
\end{abstract}

\pacs{Valid PACS appear here}
\maketitle

\section{Transient absorption spectroscopy}
The absorption cross-section can be computed as~\cite[p.~411]{tannor2007}
\begin{equation}
\label{eq:abs_cross_section}
  \sigma(\omega) = 
    -\frac{4\pi \omega}{c}
    \frac{\text{Im}[\tilde{P}^*(\omega) \tilde{E}(\omega)]}{|\tilde{E}(\omega)|^2},
\end{equation}
where $\omega$ is the photon energy, $c$ is the speed of light in vacuum, $\tilde{E}(\omega)$ is the Fourier transform of the electric field $\tilde{E}(\omega) = \mathcal{F}[E(t)]$, and $\tilde{P}(\omega)$ is the Fourier transform of the polarization function
\begin{equation}
	P(t) = \langle \Psi(t) | \hat{\mu} | \Psi(t) \rangle,
\end{equation}
where $\hat{\mu}$ is the electric dipole operator, and $|\Psi(t)\rangle$ is the time-dependent wavefunction. In order to calculate the field-induced polarization response, we need to solve the time-dependent Schr\"odinger equation
\begin{equation}
  i \frac{d |\Psi(t)\rangle}{d t} = \hat{H}(t) |\Psi(t)\rangle,
\end{equation}
with the Hamilton operator 
\begin{equation}
	\hat{H}(t) = \hat{H}_0 - \hat{\mu} E(t),
\end{equation}
describing a system under the influence of an external field. Here, the scalar dipole operator $\hat{\mu} = \hat{\vec{\mu}} \cdot \vec{\epsilon}$ denotes the projection of the dipole vector operator $\hat{\vec{\mu}}$ onto the unit polarization vector $\vec{\epsilon}$ of an external field $\vec{E}(t) = \vec{\epsilon}\ E(t)$.

To evaluate the polarization $P(t)$, we employ the time-dependent perturbation theory, in which the state at time $t$ is computed as
\begin{equation}
\label{eq:psi_ptb_expans}
 | \Psi(t) \rangle = 
    \hat{U}(t) | \Psi(0) \rangle +
    i \int_{-\infty}^t dt' \hat{U}(t-t') \hat{\mu} E(t') \hat{U}(t') | \Psi(0) \rangle + \cdots,
\end{equation}
where $\hat{U}(t)$ is the field-free evolution operator for the Hamiltonian $\hat{H}_0$ and $|\Psi(0)\rangle$ is the initial state of the system at time zero.

Using Eq.~(\ref{eq:psi_ptb_expans}), we can expand the polarization $P(t)$ in powers of the electric field as
\begin{equation} \label{eq:P_exp}
	P(t) = P^{(0)}(t) + P^{(1)}(t) + \cdots,
\end{equation}
where the zeroth-order term $P^{(0)}(t)$ describes the harmonic emission, resulting from oscillations of the dipole moment in time (see, e.g., Ref.~\cite{kuleff2011}), and
\begin{equation}
\label{eq:P_expansion}
	P^{(1)}(t) = i \langle \Psi(0) | \hat{U}^{\dagger}(t) \hat{\mu} \int_{-\infty}^t dt' \hat{U}(t-t') \hat{\mu} E(t') \hat{U}(t') | \Psi(0) \rangle + \text{c.c.},
\end{equation}
is the linear polarization response with respect to the applied electric field. Importantly, in an isotropic medium even-order polarization terms (i.e. $P^{(0)}(t)$, $P^{(2)}(t)$, etc.) do not survive orientational averaging which will be employed later and thus can be ignored~\cite{tannor2007}. The higher-order terms of the polarization expansion, Eq.~(\ref{eq:P_exp}), are responsible for various processes that enter in a broad variety of non-linear spectroscopy techniques~\cite{mukamel1995book}, which are beyond the scope of the present study and thus will be neglected. Henceforth, hereafter we will concentrate on the linear polarization response function, Eq.~(\ref{eq:P_expansion}), which describes absorption and stimulated emission driven by the external electric field $E(t)$.

Let us assume for the moment that the eigenvalues $\varepsilon_k$ and eigenvectors $|\Psi_k \rangle$ of the operator $\hat{H}_0$ are known. Since
\begin{equation}
\label{eq:eigensystem_H0}
  \hat{H}_0 |\Psi_k \rangle = \varepsilon_k |\Psi_k \rangle,
\end{equation}
the action of the evolution operator $\hat{U}(t)$ on the initial state $|\Psi(0)\rangle$ can be represented as
\begin{equation}
\label{eq:U_on_psi0}
 \hat{U}(t) | \Psi(0) \rangle = 
    \sum_k c_k e^{-i \varepsilon_k t} | \Psi_k \rangle,
\end{equation}
where $c_k = \langle \Psi_k | \Psi(0) \rangle$ are the expansion coefficients of the initial state in a basis of $\hat{H}_0$ eigenstates. Using the explicit form of the evolution operator acting on the initial state in Eq.~(\ref{eq:U_on_psi0}), we can write the first-order polarization correction as
\begin{equation}
  P^{(1)}(t) = i \int_{-\infty}^t dt' E(t')
    \sum_k \sum_j c^*_k c_j e^{i \varepsilon_k t} e^{-i \varepsilon_j t'} 
    \langle \Psi_k | \hat{\mu} \hat{U}(t-t') \hat{\mu} | \Psi_j \rangle + \text{c.c.}
\end{equation}
To evaluate the matrix elements $\langle \Psi_k | \hat{\mu} \hat{U}(t-t') \hat{\mu} | \Psi_j \rangle$, we insert the resolution of identity $\sum_f |\Psi_f \rangle \langle \Psi_f | \equiv 1$ between $\hat{U}(t-t')$ and $\hat{\mu}$ operators, thus obtaining
\begin{equation}
\label{eq:P_t_full}
  P^{(1)}(t) = i \sum_k \sum_j c^*_k c_j 
    \sum_f \langle \Psi_k | \hat{\mu} | \Psi_f \rangle
           \langle \Psi_f | \hat{\mu} | \Psi_j \rangle
	\underbrace{   
    		\int_{-\infty}^t dt' E(t')
   			e^{i \varepsilon_k t} e^{-i \varepsilon_j t'}
    			e^{-i \varepsilon_f (t-t')}}_{I(t)} + \text{c.c.}
\end{equation}
By reorganizing $\varepsilon$ energies on the right-hand side of Eq.~(\ref{eq:P_t_full}), the time integral $I(t)$ can be represented in the form of a convolution
\begin{equation}
    I(t) = \int_{-\infty}^t dt' E(t')
    		e^{i (\varepsilon_k - \varepsilon_j) t'}
    		e^{-i (\varepsilon_f - \varepsilon_k) (t-t')} = 
    		\int_{-\infty}^{\infty} dt' E(t')
    		e^{i (\varepsilon_k - \varepsilon_j) t'}
    		e^{-i (\varepsilon_f - \varepsilon_k) (t-t')} \theta(t-t'),
\end{equation}
where $\theta(t-t')$ is the Heaviside step function. The convolution of two functions in the time domain gives a simple product in the frequency domain (Fourier convolution theorem)
\begin{equation}
\label{eq:integral_w}
	\tilde{I}(\omega) = \sqrt{2\pi}
		\mathcal{F}[E(t) e^{i (\varepsilon_k - \varepsilon_j) t}]
		\mathcal{F}[e^{-i (\varepsilon_f - \varepsilon_k) t} \theta(t)],
\end{equation}
where both terms in the product can be evaluated analytically:
\begin{equation}
\label{eq:FT_1}
	\mathcal{F}[E(t) e^{i (\varepsilon_k - \varepsilon_j) t}] = 
		\tilde{E}(\omega - \epsilon_k + \epsilon_j),
\end{equation}
and
\begin{equation}
\label{eq:FT_2}
	\mathcal{F}[e^{-i (\varepsilon_f - \varepsilon_k) t} \theta(t)] = 
		\frac{1}{\sqrt{2\pi}} \frac{-i}{\epsilon_f - \epsilon_k - \omega}.
\end{equation}
Importantly, the energy difference $\epsilon_f - \epsilon_k$ between states coupled by the electric field has to be taken complex with a negative imaginary part in order to converge the Fourier transform in Eq.~(\ref{eq:FT_2}). From a physical point of view, complex values of energy levels can be associated with a finite lifetime of final states which will result in the broadening of the absorption lines in the spectrum.

Approximating the laser field by a delta function centered at $t=\tau$, i.e., $E(t) = E_0 \delta (t-\tau)$, the Fourier transform $\tilde{E}(\omega)$ of the electric field yields
\begin{equation}
	\tilde{E}(\omega) = 
		\frac{1}{\sqrt{2\pi}} E_0 e^{i \omega \tau},
\end{equation}
while the Fourier transform in Eq.~(\ref{eq:FT_1}) becomes
\begin{equation}
	\tilde{E}(\omega - \epsilon_k + \epsilon_j) = 
		\frac{1}{\sqrt{2\pi}} E_0 e^{i (\omega + \epsilon_k - \epsilon_j) \tau}.
\end{equation}
Substituting the above presented relations into Eq.~(\ref{eq:abs_cross_section}), we obtain the following expression for the absorption cross-section (see also Ref.~\cite{santra2011})
\begin{equation}
\label{eq:abs_cs_full}
  \sigma (\omega,\tau) = \frac{4\pi\omega}{c} 
  	\text{Im} \sum_k \sum_j 
    c^*_k e^{ i \varepsilon_k \tau}
    c_j e^{-i \varepsilon_j \tau}
    \sum_f \langle \Psi_k | \hat{\mu} | \Psi_f \rangle
           \langle \Psi_f | \hat{\mu} | \Psi_j \rangle
    \left(
      \frac{1}{\varepsilon_f - \varepsilon_k - \omega}
    + \frac{1}{\varepsilon_f^* - \varepsilon_j + \omega}
    \right).
\end{equation}

So far, we did not specify the physical nature of the Hamiltonian $\hat{H}_0$ and the corresponding eigenstates and eigenenergies, Eq.~(\ref{eq:eigensystem_H0}). Let us consider a particular case, where the operator $\hat{H}_0$ is the full molecular Hamiltonian
\begin{equation}
	\hat{H}_0 = \hat{T}_n + \hat{H}_e,
\end{equation}
where $\hat{T}_n$ is the kinetic energy operator of the nuclei and $\hat{H}_e$ denotes the electronic Hamiltonian. Using the well-known Born--Oppenheimer formalism, i.e. solving the eigenvalue equation $\hat{H}_e \Phi_I(\mathbf{r},\mathbf{R}) = E_I (\mathbf{R}) \Phi_I(\mathbf{r},\mathbf{R})$, we can replace the summation over molecular states in Eq.~(\ref{eq:abs_cs_full}) with a summation over the electronic states:
\begin{equation}
  \sum_j c_j e^{-i \varepsilon_j \tau} |\Psi_j \rangle =
  \sum_J \chi_J(\mathbf{R},\tau) |\Phi_J \rangle,
\end{equation}
where $\chi_J(\mathbf{R},\tau)$ are nuclear wave packets propagating on the corresponding potential energy surfaces $E_J (\mathbf{R})$. Furthermore, we assume that vibrationally-resolved absorption lines in Eq.~(\ref{eq:abs_cs_full}) can be approximated with good accuracy by the averaged transitions between initial $\epsilon_j \approx E_J$ and final $\epsilon_f \approx \tilde{E}_F = E_F - i \frac{\Gamma}{2}$ electronic energy levels, energetically broadened by the parameter $\Gamma$. The particular value of $\Gamma$ has to be taken such that it accounts for the core-hole lifetimes ($\sim$160 meV for the O$_{\text{1s}}$~\cite{sankari2003} and $\sim$100 meV for C$_{\text{1s}}$~\cite{carroll2000}) and the experimental resolution in ATAS (typically on the order of 0.2-0.3 eV at the X-ray energy range employed in the present work (see, e.g., Ref.~\cite{geneaux2019})). The final form of the cross-section for electronically resolved linear absorption spectroscopy reads
\begin{equation}
\label{eq:cs_final}
\begin{split}
  \sigma (\omega,\tau) = \frac{4\pi\omega}{c} 
  & \text{Im} \sum_I \sum_J 
    \langle \chi_I(\mathbf{R},\tau) | \chi_J(\mathbf{R},\tau) \rangle_{\mathbf{R}}
    \sum_F \langle \Phi_I | \hat{\mu} | \Phi_F \rangle
           \langle \Phi_F | \hat{\mu} | \Phi_J \rangle \\ & \times \left(
      \frac{1}{\tilde{E}_F - E_I - \omega}
    + \frac{1}{\tilde{E}^*_F - E_J + \omega}
    \right),
\end{split}
\end{equation}
where the dynamics of initially populated states is driven by the electronic coherence terms $\langle \chi_I(\mathbf{R},\tau) | \chi_J(\mathbf{R},\tau) \rangle_{\mathbf{R}}$, and the absorption lines occur between electronic states coupled by electric dipole transitions.

\section{Transition properties between excited states}
To calculate transition properties between states computed at different levels of \textit{ab initio} electronic structure theory, we employed a general technique exploiting the configuration-interaction (CI)-like structure of the electronic wavefunctions. The starting point is a CI expansion of an $N$-electron wavefunction in terms of various excitations taking place from the reference state $|\Psi_0\rangle$
\begin{equation}
\label{eq:CI_expansion}
	|\Phi_I \rangle = \sum_p a^I_p \hat{C}_p | \Psi_0 \rangle,
\end{equation}
where $a^I_p$ are, in general, complex expansion coefficients, and $\hat{C}_p$ is a formal representation of an operator describing excitations in terms of creation $\hat{c}^{\dagger}$ and annihilation $\hat{c}$ operators acting on the reference state. The representation of the wavefunction in the form of Eq.~(\ref{eq:CI_expansion}) is general and appears in many methods of electronic structure theory such as CI of various types~\cite{helgaker2000}, equation-of-motion coupled-clusters (EOM-CC) family of methods~\cite{krylov2008}, and algebraic diagrammatic construction (ADC) scheme~\cite{schirmer2018book}, to name a few.

A particular excitation $p$ in Eq.~(\ref{eq:CI_expansion}) represents a manipulation with electrons occupying certain orbitals of the system which can be conveniently described in terms of Slater determinants. Mathematically, the Slater determinant of the reference state $|\Psi_0\rangle$ is defined as an antisymmetrized product of orthonormal one-electron molecular spin-orbitals $\{\varphi_1,\varphi_2,...,\varphi_N\}$
\begin{equation}
\label{eq:slater_determinant}
\begin{aligned}
	|\Psi_0\rangle =
  		\frac{1}{\sqrt{N!}}
  		\begin{vmatrix} \varphi_1(\mathbf{x}_1) & \varphi_2(\mathbf{x}_1) & \cdots & \varphi_N(\mathbf{x}_1) \\
                        \varphi_1(\mathbf{x}_2) & \varphi_2(\mathbf{x}_2) & \cdots & \varphi_N(\mathbf{x}_2) \\
                        \vdots & \vdots & \ddots & \vdots \\
                        \varphi_1(\mathbf{x}_N) & \varphi_2(\mathbf{x}_N) & \cdots & \varphi_N(\mathbf{x}_N)
  		\end{vmatrix} \equiv
  		| \varphi_1,\varphi_2, ..., \varphi_N \rangle,
\end{aligned}
\end{equation}
where $\mathbf{x}_i$ denotes the spin and spatial coordinates of the $i$-th electron. Let us consider an excitation $\hat{C}_p = \hat{c}^{\dagger}_a \hat{c}_k$ generating one-hole--one-particle (1$h$-1$p$) configurations: an electron is removed from an occupied spin-orbital $\varphi_k$ and an electron is added to a virtual spin-orbital $\varphi_a$
\begin{equation}
	|\Psi_p\rangle = \hat{C}_p |\Psi_0\rangle =
	\hat{c}^{\dagger}_a \hat{c}_k | \varphi_1 ... \varphi_k ... \varphi_N \rangle 	=
	| \varphi_1 ... \varphi_a ... \varphi_N \rangle,
\end{equation}
where in the singly excited Slater determinant $|\Psi_p\rangle$ the spin-orbital $\varphi_k$ is replaced by another spin-orbital $\varphi_a$. Similarly, any kind of electronic excitation $p$ can be represented by adding or removing spin-orbitals from the reference Slater determinant, Eq.~(\ref{eq:slater_determinant}).

Representing the wavefunctions from Eq.~(\ref{eq:CI_expansion}) in terms of Slater determinants allows us to express the matrix elements of a one-electron operator $\hat{F}$ between excited states $I$ and $J$ as
\begin{equation}
\label{eq:F_SlatDeterm}
	\langle \Phi_I|\hat{F}|\Phi_J \rangle = \sum_p \sum_q (a^I_p)^* a^J_q \langle \Psi_p |\hat{F}| \Psi_q \rangle,
\end{equation}
being proportional to the sum of all matrix elements $\langle \Psi_p |\hat{F}| \Psi_q \rangle$ between the Slater determinants $p$ and $q$. These matrix elements can be expressed in terms of integrals in the spin-orbital space using Slater--Condon rules~\cite{slater1929,condon1930} (see also Ref.~\cite{helgaker2000} for a general review). For a one-electron operator, these rules are:
\begin{equation}
\label{eq:SCR_1}
	\langle \varphi_1,\varphi_2, ..., \varphi_N | \hat{F} | \varphi_1,\varphi_2, ..., \varphi_N \rangle =
	\sum_{i=1}^N \langle \varphi_i | \hat{F} | \varphi_i \rangle,
\end{equation}
i.e., the expectation value of the operator $\hat{F}$ between identical Slater determinants is obtained by summing up matrix elements $\langle \varphi_i | \hat{F} | \varphi_i \rangle$ between spin-orbitals;
\begin{equation}
\label{eq:SCR_2}
	\langle \varphi_1 ... \varphi_k ... \varphi_N | \hat{F} | \varphi_1 ... \varphi_a ... \varphi_N \rangle =
	\langle \varphi_k | \hat{F} | \varphi_a \rangle,
\end{equation}
i.e. the matrix element between two Slater determinants differing by a single orbital is equal to the integral between corresponding spin-orbitals; and 
\begin{equation}
\label{eq:SCR_3}
	\langle \varphi_1 ... \varphi_k ... \varphi_j ... \varphi_N | 
		\hat{F} | \varphi_1 ... \varphi_a ... \varphi_b ... \varphi_N \rangle = 0,
\end{equation}
when the Slater determinants differ by two or more spin-orbitals. Therefore, computation of matrix elements of a one-electron operator between excited states boils down to evaluating transition integrals between spin-orbitals of the system.

Importantly, a particular form of the spin-orbital basis depends on the nature of the reference state $|\Psi_0\rangle$ from which the excitations are taking place. A simplest case is a standard CI scheme where the reference state is the Hartree--Fock wavefunction, i.e. the spin-orbitals are obtained by the diagonalization of the Fock matrix, and the state is a single Slater determinant, Eq.~(\ref{eq:slater_determinant}). In more sophisticated methods such as EOM-CC or ADC, the reference state is correlated and thus additional transformations of the spin-orbital basis are required. In the case of ADC scheme used in the present work, the reference state is constructed by applying the Rayleigh--Schr\"odinger perturbation theory to the Hartree--Fock wavefunction and supposing a standard M{\o}ller--Plesset partitioning of the Hamiltonian~\cite{schirmer2018book}. Since the resulting expressions, Eqs.~(\ref{eq:F_SlatDeterm}) to~(\ref{eq:SCR_3}), for evaluating the one-particle operator $\hat{F}$ do not contain the spin-orbitals themselves but only the matrix elements of the corresponding operator, the application of the perturbation theory to the reference state is equivalent to the expansion of the operator $\hat{F}$ in a perturbation series
\begin{equation}
	\mathbf{F} = \mathbf{F}^{(0)} + \mathbf{F}^{(1)} + \mathbf{F}^{(2)} + \cdots,
\end{equation}
where $\mathbf{F}$ denotes representation of the operator $\hat{F}$ in a basis of spin-orbitals. The derivation of the explicit expressions for computing the corresponding contributions $\hat{F}^{(i)}$ is beyond the scope of the present work. However, we would like to note that the required matrix elements $F_{kl}$ of the one-particle operator up to the second order of perturbations can be straightforwardly obtained by combining the expressions for the hole/hole and particle/hole terms given in appendices of Refs.~\cite{schirmer2004,trofimov2005}.

\section{Orientational averaging of the transient absorption spectrum}
Expression~(\ref{eq:cs_final}) is derived for a particular orientation of a molecule with respect to the applied field. In the situation when the molecules are oriented randomly in space, the cross-section can be obtained by averaging over all possible orientations with respect to the laser polarization (see also Ref.~\cite{begusic2018_HT} and references therein):
\begin{equation}
\label{eq:cs_aver}
\begin{split}
	\sigma (\omega,\tau) = & \int_{0}^{2\pi} \int_{0}^{\pi} 
		\vec{\epsilon}(\theta,\varphi)^{T} \cdot \bm{\sigma}_{\vec{\mu},\vec{\mu}}(\omega,\tau)
			\cdot \vec{\epsilon}(\theta,\varphi) d \theta d \varphi \\
			= & \frac{1}{3} \left( 
				\sigma_{\mu_x,\mu_x}(\omega,\tau) + \sigma_{\mu_y,\mu_y}(\omega,\tau) + \sigma_{\mu_z,\mu_z}(\omega,\tau)
			\right),
\end{split}			
\end{equation}
where $\vec{\epsilon}(\theta,\varphi)^T = \left(\sin{(\theta)} \cos{(\varphi)},\ \sin{(\theta)} \sin{(\varphi)},\ \cos{(\theta)}\right)$ is the radial unit vector, $\bm{\sigma}_{\vec{\mu},\vec{\mu}}$ is a tensor obtained by combining different orthogonal projections of the dipole operator $\hat{\vec{\mu}}$ in the corresponding transitions $\langle \Phi_I | \hat{\mu} | \Phi_F \rangle$ and $\langle \Phi_F | \hat{\mu} | \Phi_J \rangle$ present in Eq.~(\ref{eq:cs_final}), and $\sigma_{\mu_x,\mu_x}(\omega,\tau)$, $\sigma_{\mu_y,\mu_y}(\omega,\tau)$, $\sigma_{\mu_z,\mu_z}(\omega,\tau)$ are absorption cross-sections computed for three mutually orthogonal orientations of the molecule.

\begin{figure}
\captionsetup[subfigure]{labelformat=empty}
\subfloat{
	\includegraphics[width=0.3\textwidth]{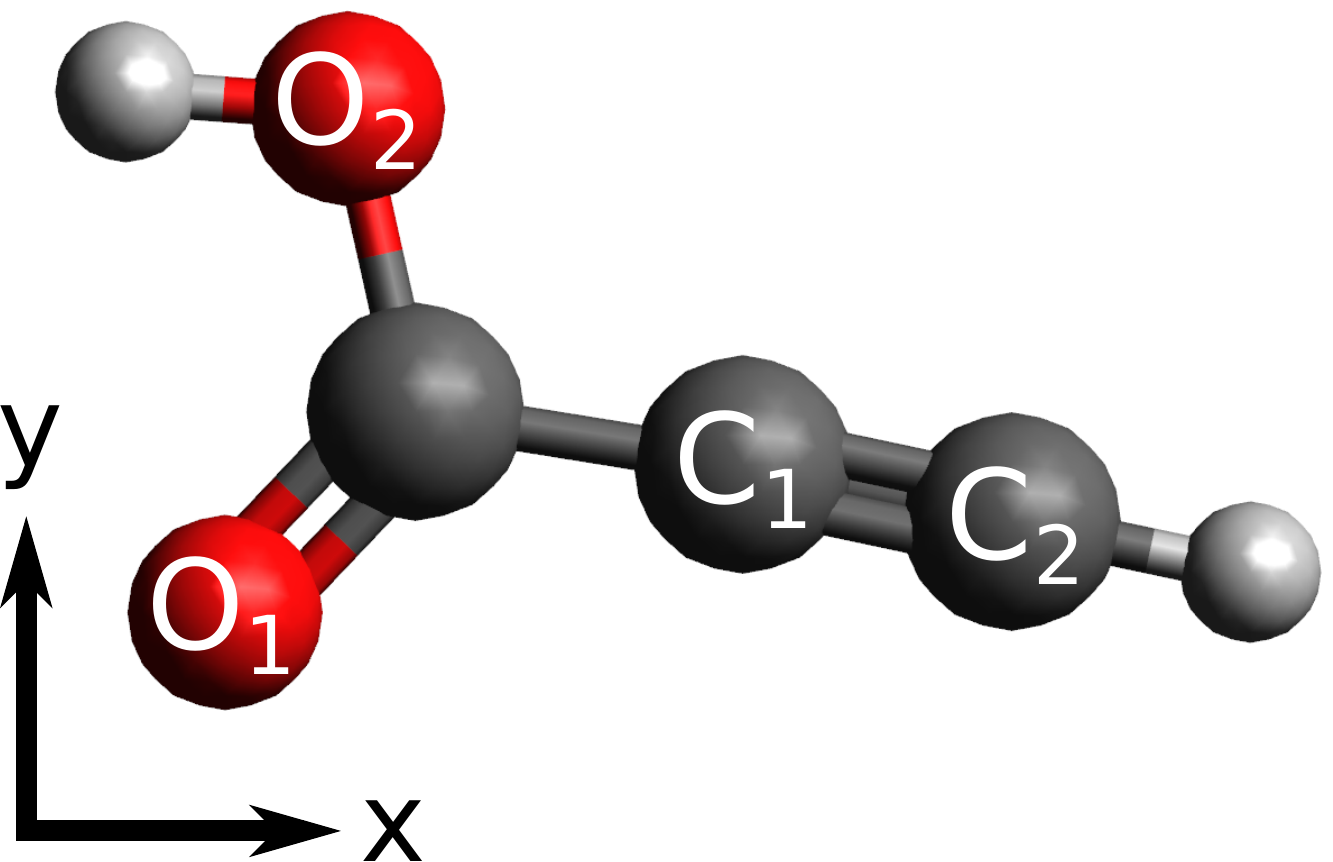}
}\\
\subfloat[(a) Absorption in $x$ direction.]{
	\includegraphics[width=0.45\textwidth]{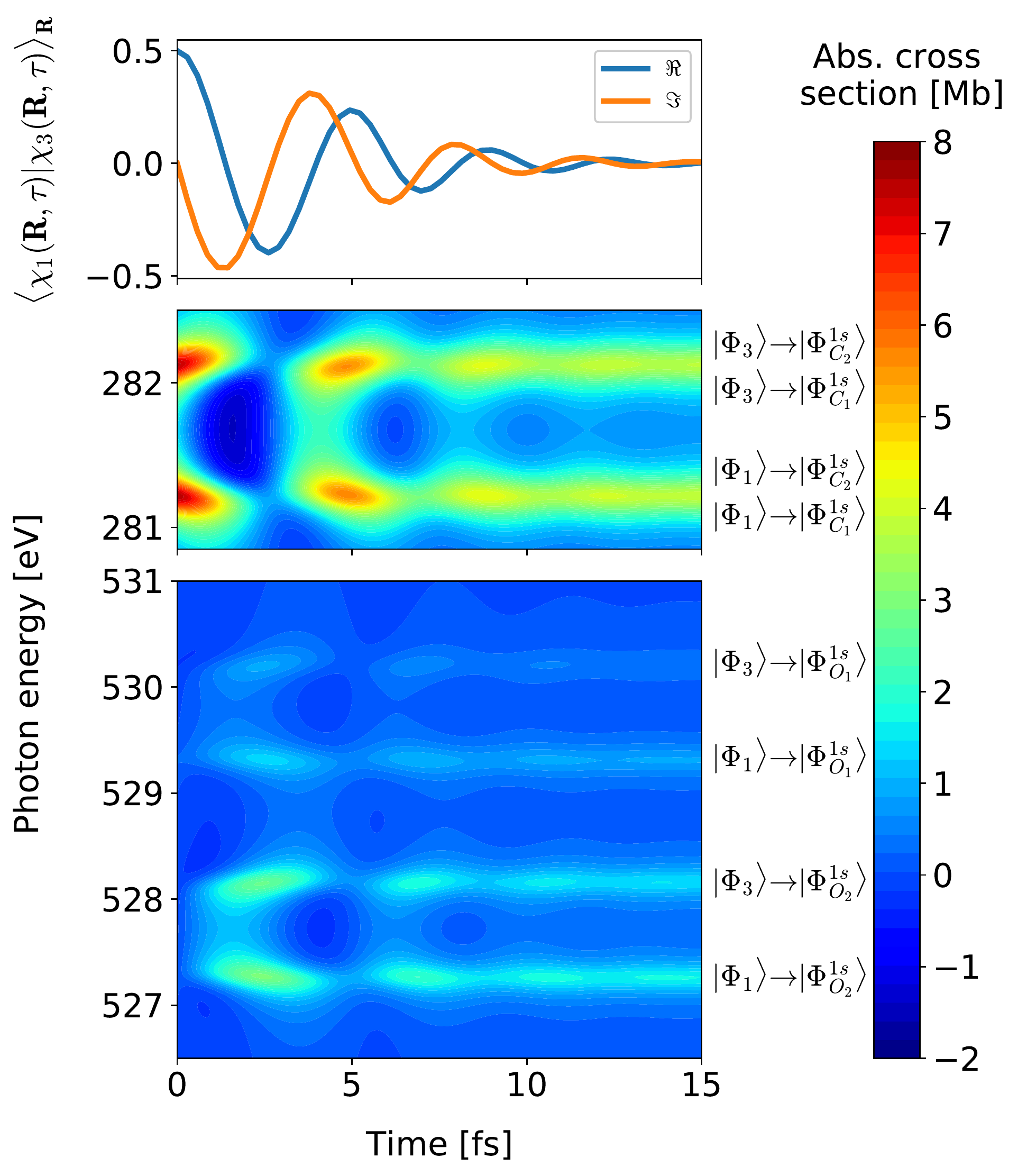}
}
\subfloat[(b) Absorption in $y$ direction.]{
	\includegraphics[width=0.45\textwidth]{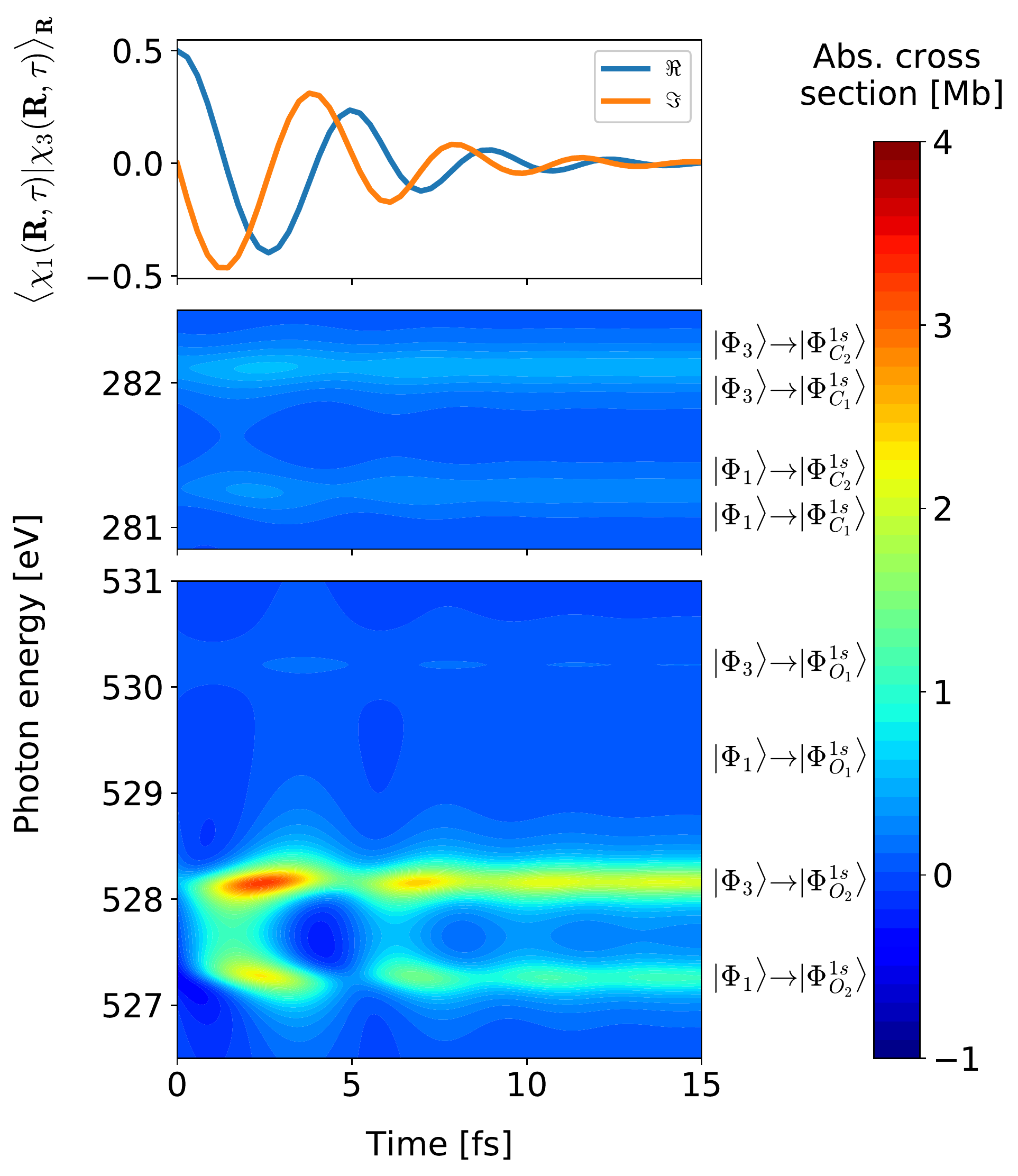}
}
\caption{Time-resolved absorption cross-section as a function of the photon energy and time delay along two orthogonal directions of propiolic acid molecule: $x$ (left) and $y$ (right). Top panels: Electronic coherence measured by the time-dependent overlap $\langle \chi_1 (\mathbf{R},\tau) | \chi_3 (\mathbf{R},\tau) \rangle_{\mathbf{R}}$ of the nuclear wave packets propagated in the first and third cationic states of propiolic acid after the removal of an electron from the HOMO. Middle panel: The absorption cross-section plotted for the energy window corresponding to transitions between initially populated valence ionic states and the core states resulting from ionization out of $1s$ orbitals of carbon atoms forming the triple bond. Bottom panel: The absorption cross-section plotted for the energy range covering transitions to core ionic states of oxygen atoms. The ionic states involved in the corresponding transitions are also shown.}
\label{fig:ATAS_xyz}
\end{figure}

The absorption cross-sections for different orientations of the propiolic acid molecule with respect to the polarization of the applied field are shown in Fig.~\ref{fig:ATAS_xyz}. Due to the planar geometry of the molecule and the fact that the initial superposition of the electronic states is prepared by ionizing an electron from the HOMO belonging to A$'$ symmetry (see Fig.~1 of the main text), the electric dipole transitions oriented perpendicularly to the molecular plane vanish leading thus to complete transmission of the electric field polarized in this direction. In contrast, a laser pulse polarized in the molecular plane is absorbed in such a way that the maximum of the absorption from initially created superposition of valence states to core ionic states coincides with the localization of the charge density in the vicinity of a particular atom. As one can see, the absorption on the triple-bond carbon atoms takes place almost exclusively along $x$ direction of the molecule while only a weak signal appears along $y$ direction. Indeed, the orbitals depicted in Fig.~1 of the main text suggest that the positive and negative contributions of the HOMO (shown in red and blue colors in Fig.~1 of the main text) in the vicinity of the triple-bond along $y$ direction of the molecule cancel each other out leading thus to nearly zero dipole moments connecting the HOMO with the core orbitals of the carbon atoms in this direction. It is also seen from the same figure that the HOMO-2 is mostly localized in the vicinity of the carboxyl group which results in the weak, yet distinguishable, dipole couplings with the core orbitals of the carbon atoms along both directions. The stronger dipole transitions to the carbon orbitals along $y$ direction from the HOMO-2 in comparison with those from the HOMO is reflected in the delay of the maximum of the absorption signal depicted in the right panel of Fig.~1 of SM. Following similar logic, we can see from Fig.~1 of the main text that the dipole transitions to the oxygen core orbitals are expected to be stronger from the HOMO-2 in comparison with those from the HOMO. Indeed, it is seen from Fig.~1 of SM that the maximum of the absorption involving oxygen core orbitals is reached when the charge initially created in the HOMO migrates to the HOMO-2. The averaged absorption cross-section obtained through Eq.~(\ref{eq:cs_aver}) is shown in Fig.~2 of the main text.

\section{Averaging of the transient absorption spectrum with respect to the relative contributions of ionic states to the initial wave packet}
The sudden ionization limit employed in the present work assumes that the ionization is performed with a short high-energy laser pulse. Due to a large bandwidth of such a pulse, one can expect that the initial wave packet will be prepared by a coherent superposition of multiple ionic states covered by the applied pulse. Importantly, the large energy gap between the four valence ionic states of the propiolic acid shown in Fig.~1 of the main text and the ionic states lying higher  in energy makes it possible to disentangle the high-energy contributions from the transient absorption spectrum presented in this work. Indeed, the electronic coherences between the lowest four cationic states and the higher states will have oscillation periods well below a femtosecond and can be removed from the spectrum, for example, by Fourier analysis of the ATAS signal. Furthermore, the second and the fourth ionic states shown in Fig.~1 of the main text are composed almost exclusively from the one-hole configurations that overlap neither with each other, nor with the configurations building the first and third cationic states of interest. The lack of spatial overlap between the states in the created wave packet leads to trivial phase dynamics, while observable properties of the system, such as the charge density, will remain stationary in time. Therefore, the absorption lines corresponding to transitions from the second and fourth ionic states to the core ionic states will not oscillate and thus can be filtered out from the spectrum.

\begin{figure}
\captionsetup[subfigure]{labelformat=empty}
\subfloat[(a) Only first cationic state is populated.]{
	\includegraphics[height=5.5cm]{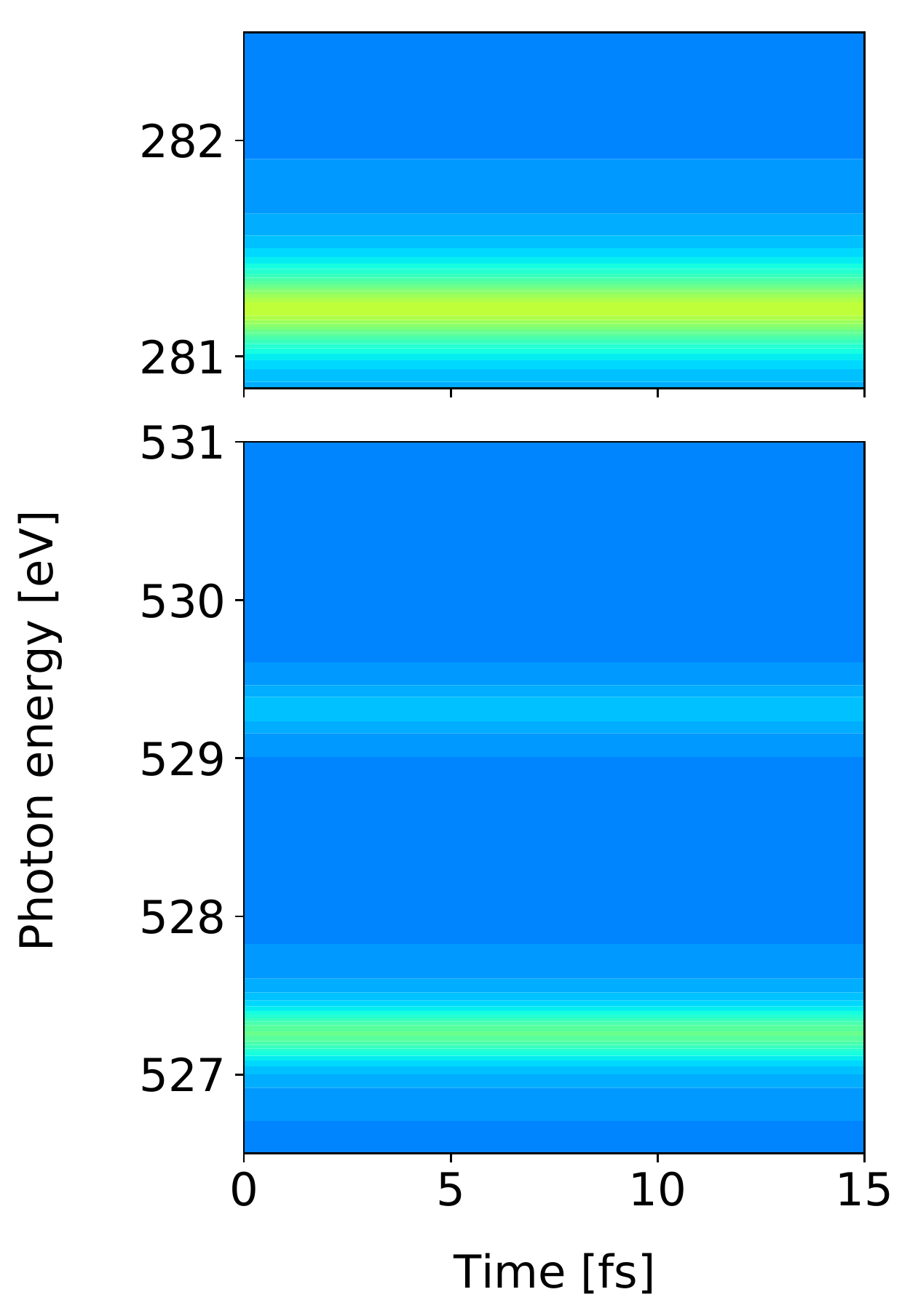}
}
\subfloat[(b) Only third cationic state is populated.]{
	\includegraphics[height=5.5cm]{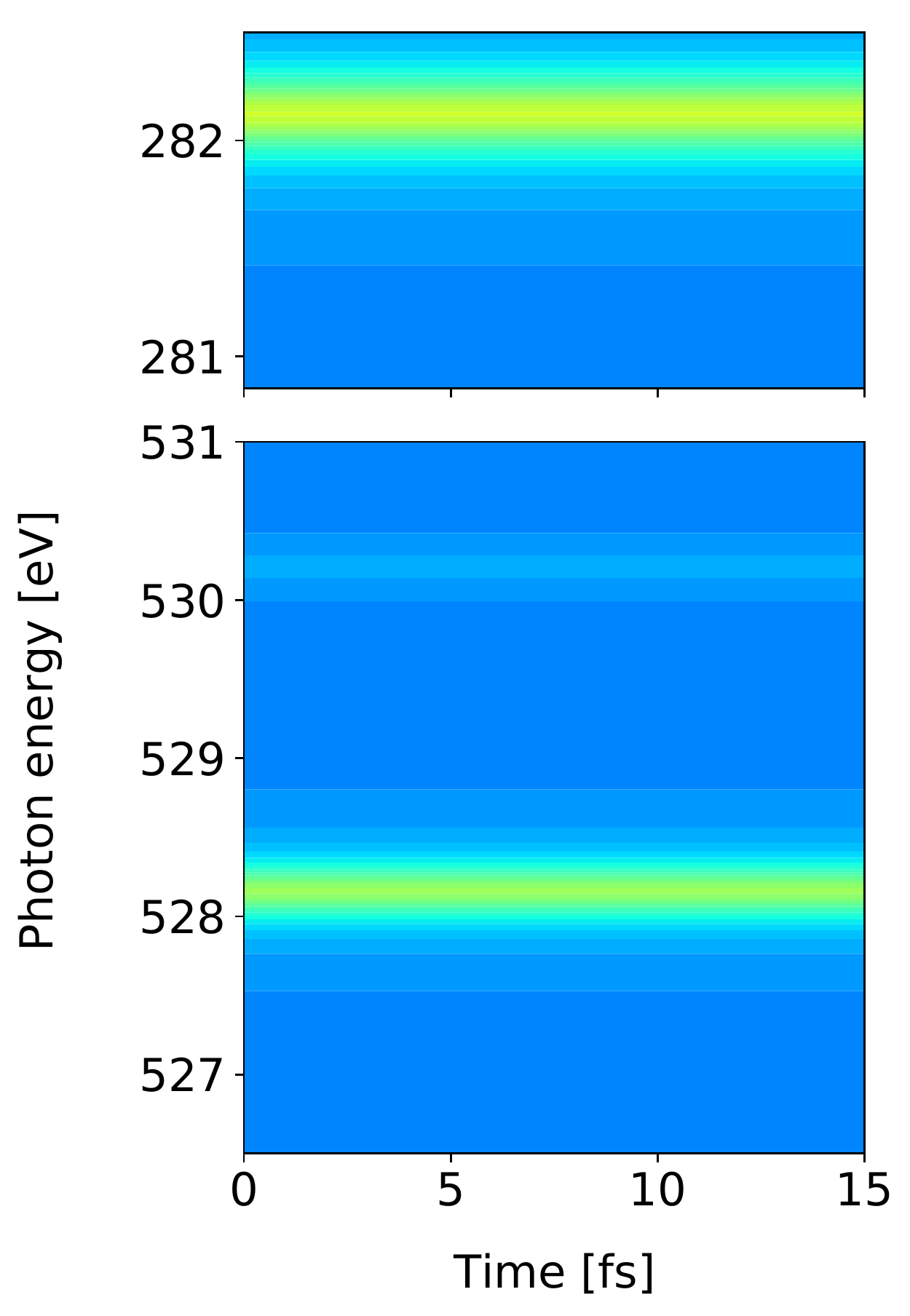}
}
\subfloat[(c) Averaging over various populations of the first and the third cationic states.]{
	\includegraphics[height=5.5cm]{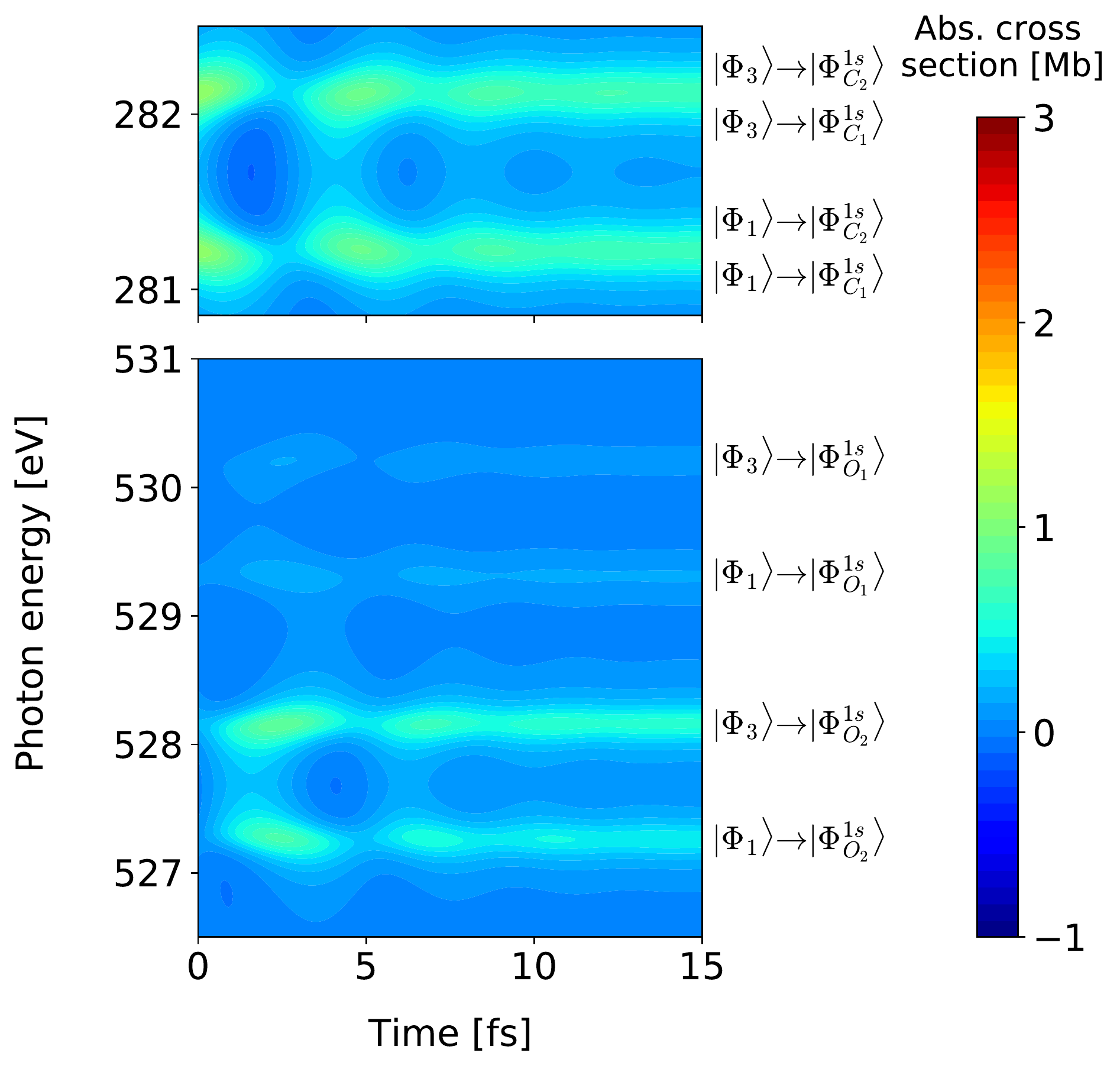}
}
\caption{Time-resolved absorption cross-section as a function of the photon energy and time delay obtained for different relative contributions of ionic states to the initial wave packet created after ionization of the propiolic acid. Left panel: Only the first cationic state is populated. Middle panel: Only the third cationic state is populated. Right panel: The averaged spectrum obtained by varying the relative contributions of the first and the third ionic states in the initial wave packet. Top and bottom panels in each case represent energy windows corresponding to transitions between initially populated valence ionic states and the core states resulting from ionization out of $1s$ orbitals of carbon and oxygen atoms, respectively.}
\label{fig:ATAS_wp_aver}
\end{figure}

The instability of laser pulse parameters, as well as the random orientation of molecules in space might lead to variations of the populations of the ionic states in the initial superposition. The absorption cross-sections for different relative contributions of ionic states to the initial wave packet created after ionization of the propiolic acid are shown in Fig.~\ref{fig:ATAS_wp_aver}. Left and middle panels of Fig.~\ref{fig:ATAS_wp_aver} illustrate the two limiting situations when the initial state is prepared exclusively from the first or the third cationic state, respectively. It is seen that in both cases the absorption cross-sections remain stationary in time, whereby only the lines corresponding to transitions from the first or the third valence state to the core ones are present. Right panel of Fig.~\ref{fig:ATAS_wp_aver} demonstrates the averaged absorption cross-section obtained by varying the relative contributions of the first and the third ionic states to the initial wave packet, i.e., $\chi_1 (\mathbf{R},0) = a \chi_0 (\mathbf{R})$ and $\chi_3 (\mathbf{R},0) = \sqrt{1-a^2} \chi_0 (\mathbf{R})$, where $\chi_0 (\mathbf{R})$ is the neutral ground state of the molecule. The resulting spectrum is obtained by arithmetical averaging of ATAS signals computed for eleven values of parameter $a$ taken between 0 and 1. As one can see, the ATAS signal survives the averaging and thus provides sufficient information to trace temporal and spatial aspects of the underlying electron dynamics.

\bibliographystyle{apsrev4-1}
\bibliography{ATAS_decoherence.bib}